\documentclass[a4paper]{article}

\usepackage{amsmath,graphicx}
\usepackage{color}
\usepackage{balance}
\usepackage{multicol}
\usepackage{subfigure}
\usepackage{graphicx}
\usepackage{comment}
\usepackage{INTERSPEECH2020}
\title{Improving Embedding Extraction for Speaker Verification with Ladder Network}
\name{Fei Tao, Gokhan Tur}
\address{Convsational AI, Uber AI Labs}
\email{fei.tao@hotmail.com, gokhan.tur@ieee.org}

\begin{document}

\maketitle
\thispagestyle{myheadings}
\begin{abstract}
Speaker verification is an established yet challenging task in speech processing and a very vibrant research area. Recent speaker verification (SV) systems rely on deep neural networks to extract high-level embeddings which are able to characterize the users' voices. Most of the studies have investigated on improving the discriminability of the networks to extract better embeddings for performances improvement. However, only few research focus on improving the generalization. In this paper, we propose to apply the ladder network framework in the SV systems, which combines the supervised and unsupervised learning fashions. The ladder network can make the system to have better high-level embedding by balancing the trade-off to keep/discard as much useful/useless information as possible. We evaluated the framework on two state-of-the-art SV systems, d-vector and x-vector, which can be used for different use cases. The experiments showed that the proposed approach relatively improved the performance by 10\% at most without adding parameters and augmented data.

\end{abstract}

\section{Introduction}
\label{sec:intro}

Speaker verification (SV) aims to verify a speaker using the speech signal. As an open set recognition problem, speaker verification performs the task where testing labels are not seen during training, and most probably, there is large mismatch between testing and training data (different environments, languages, and devices)~\cite{Dehak_2011_2,liu2017opensesame,Garcia-Romero_2011,variani2014deep,snyder2018x,Nagrani17,Chung18b}. A standard pipeline is extracting a speaker embedding relying on some trained model and comparing the similarity between the registered and testing embeddings. One key problem in this task is extracting a good embedding which can capture characteristics of the input voice to discriminate speakers. Recently, the \emph{deep neural network} (DNN) approaches are widely used since they were proved to be able to extract better high-level embeddings~\cite{variani2014deep,snyder2018x,Nagrani17,Chung18b}. The DNN approaches don't have presumption of the data distribution, in contrast, they can learn the data distribution in a data-driven fashion in manifold space~\cite{Hinton_2006,Hinton_2012,Tao_2018_4}.

Research has also been done to improve the capability of DNN to output better embeddings. Previous work has either focused on improving the capability of modeling the hidden representation from the input, e.g. \emph{convolutional neural network} (CNN), \emph{time-delay neural network} (TDNN), or focused on designing more discriminative loss functions for classification task during the neural networking training, e.g. \emph{siamese neural network} (SNN) and triplet loss. Only few efforts have investigated training strategies to maintain more original information in inputs. This is important, because supervised learning essentially discards input information regarding the targeting task within the DNN framework. We don't know what information will affect the performance on testing data during training in the open set recognition problem, like SV, it is necessary to keep input information as much as possible when we train the embedding extraction model in supervised learning fashion. In this paper, we propose to apply ladder network framework to the SV system training. Ladder networks combine supervised and unsupervised learning fashions together. On one hand, supervised learning makes the network discard useless information regarding the classification task (capture the speaker's characteristics); on the other hand, unsupervised learning makes the network keep useful information regarding the reconstruction task (improve the system generalization). Ladder network performs these two learning fashions simultaneously during training, so it guarantees to keep balance in the trade-off of keeping and discarding information.

The remaining of the paper is organized in this way: Section \ref{sec:revie} reviews previous related work; Section \ref{sec:appro} describes the details of our proposed SV system; Section \ref{sec:corpus} introduces the corpus we used and the features we extracted in this study; Section \ref{sec:exper} introduce our experiments and the results; at last, we will draw conclusions in Section \ref{sec:concl}.

\section{Related Work}
\label{sec:revie}
SV systems involve parts of speaker embedding extraction and classification. For classification, vector similarity \cite{variani2014deep} or \emph{Probabilistic linear discriminant analysis} (PLDA) \cite{ioffe2006probabilistic} is adopted as embedding classifier for verifying whether they are from same speaker or not. Choice of speaker embedding becomes the key problem mainly affecting the performance in the SV systems. Recently, \emph{deep learning} (DL) techniques have been emerging as a powerful model which is able to model complicated distribution in manifold space \cite{Hinton_2006, Bengio_2009}. Different types of DNN have been proposed for different purposes and proved to be successful in many areas \cite{Tao_2018_6,LeCun_1998_2,Yu_2013}. The key advantage of the DNN is that it does not have any assumption on the input distribution, and model the input data in a data-driven fashion \cite{Tao_2018_2,Tao_2017}. 

Snyder et al. used a TDNN to model the input acoustic feature \cite{snyder2018x}. They did statistics over a fixed duration to capture the temporal information and represent the speaker's embedding (named as x-vector), while Nagrani et al. \cite{Nagrani17} proposed to use CNN to capture high-level embedding from acoustic features. Chen and Salman also proposed a system which process frame by frame under SNN framework to improve the discriminability of DNN \cite{chen2011extracting}. Varani et al. \cite{variani2014deep} adopted a small footprint DNN taking contextual input and outputting speaker ID. The embedding (named as d-vector) extracted before the classification layer was used for verification. The x-vector is the state-of-the-art speaker verification system, which was designed to deal with long duration input and normally deployed on server, while the d-vector is a light weight system, which was designed to deal with short duration input (shorter than 1 sec) and can be deployed on portable devices.

Most of the previous work were focusing on either better network structure or more discriminative loss for extracting better embedding. Recently, Li et al. \cite{wan2018generalized} proposed a end-to-end fashion DNN-based SV system, which takes registered and testing audios as inputs and directly output whether the testing audio belongs to the registered speaker. Though the system does not have embedding extraction phase, the study is still following the idea that improve the network architecture for better performance.

Only few studies have been done on changing training strategy to maintain more original information in inputs. Previously, \emph{Denoising autoencoder} (DAE) was used to pretrain DNN by reconstructing the clean version of the signal from the noisy version of the signal \cite{Hinton_2006, bengio2007greedy}. DAE can keep most essential and robust information in the input in unsupervised learning fashion \cite{vincent2008extracting}. After pretraining, DNN was trained in supervised learning fashion. The trained DNN tends to discard some information in inputs, which the network learned to be useless with respect to the targeting task \cite{valpola2015neural}. Therefore, the final system will bias to the training domain, which is not wanted in SV systems since it is open set recognition problem, even though DAE is used for pre-training. An ideal embedding extractor should discard information regarding the loss for speaker classification task, meanwhile it can keep those original information as much as possible, which does not affect the speaker classification task loss. Valpola \cite{valpola2015neural} proposed ladder network framework to incorporate DAE training approach to DNN training at each layer during supervised learning. The parameters were updated from the cost of DAE reconstruction and targeting classification simultaneously \cite{rasmus2015semi}.

\section{Proposed Approach}
\label{sec:appro}

\subsection{Speaker Verification with DNN}
We adopted d-vector and x-vector frameworks as our baseline in this paper, considering large scale use case and light weight use case.

The d-vector framework \cite{variani2014deep} consisted of 4 \emph{fully connected} (FC) \emph{rectified linear units} (ReLUs) layers, with 512 neurons per layer; and 1 softmax layer on top of the FC layers. It is suitable to tackle with audios of short duration. We concatenate the historical and future feature frames to the current frame to construct the contextual input. We set 51 frames as the input window considering that it should be long enough to contain sufficient temporal information, and short enough to fit the application like wakeword detection. To perform quick training, the input window moves in non-overlap fashion. On top of the network, a classification layer is employed to perform \emph{speaker identity} (SID) task. The label is given corresponding to each input central frame (the current frame mentioned above). 

The x-vector framework \cite{snyder2018x} was a TDNN consisting of 9 layers. The first 5 layers were time-delay layer at frame level. The 6th layer was the statistics pooling layer, which aggregated the frame-level inputs and compute the statistics over segments. The 2 layers following the statistics pooling layer were fully connected layer, which were segment-level layer, and the last layer was the softmax layer for SID classification task. We adopted the parameters setting in \cite{snyder2018x} and skip the description for simplicity. The x-vector is the state-of-the-art framework for speaker verification, which is suitable for audios of long duration. It can have better performance than d-vector since it will model the temporal information from the statistics over the input segment.

We extracted the sequence of embedding from aforementioned networks, which were trained for SID. Averaging the embedding over the utterance form the final utterance-based speaker embedding. Length normalization over the embedding is then performed. For the d-vector framework, we compute cosine similarity between the registered and testing embedding, while we use PLDA as backend to compute similarity for x-vector framework. The threshold of the computed similarity can be determined by finding the \emph{equal error rate} (EER) point.

\subsection{Ladder Network}
\begin{figure}[tb]
\centering
\subfigure
{
 \includegraphics[width=1.10\columnwidth,height=60mm]{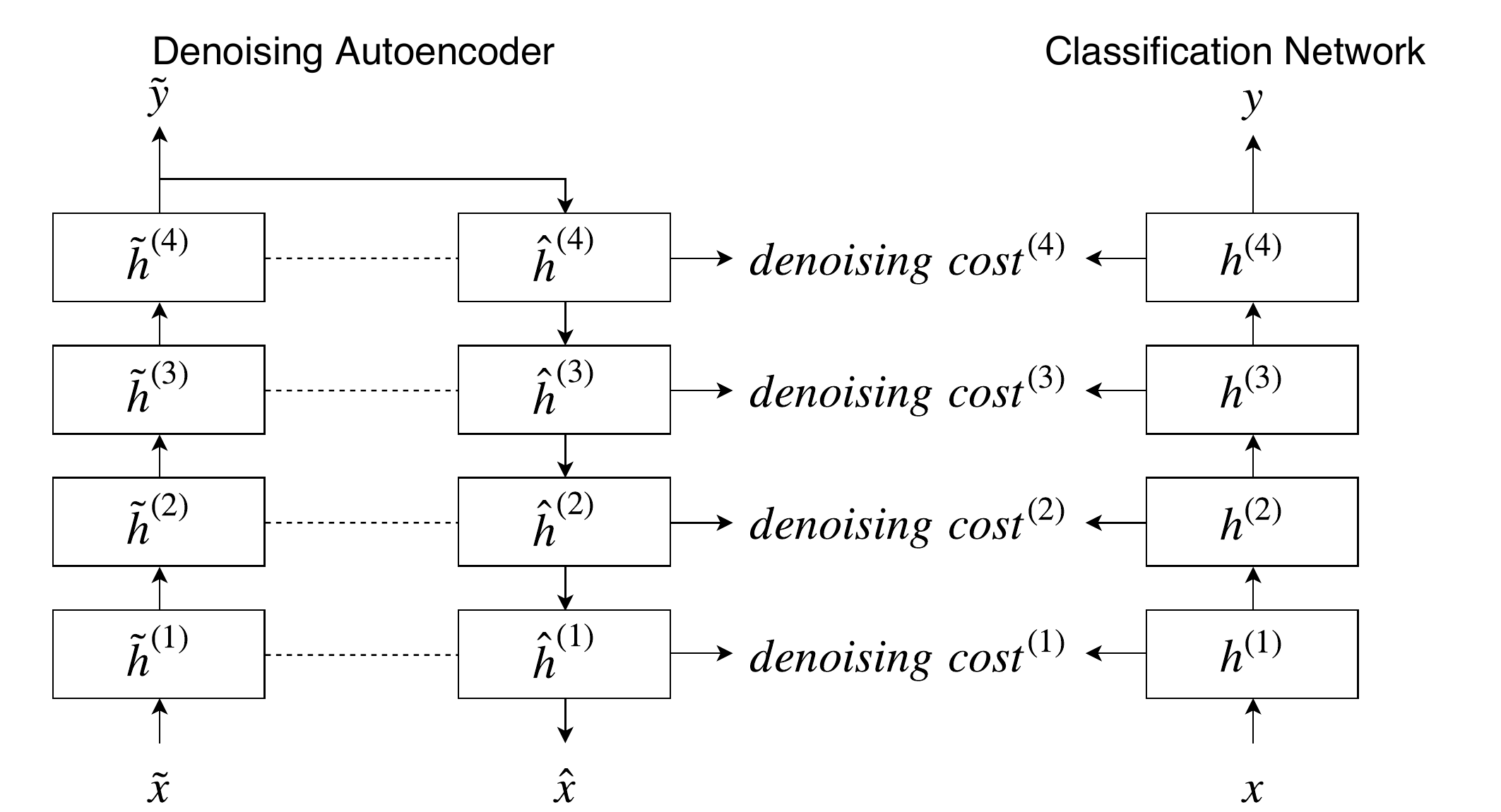}
}
\caption{The diagram of the ladder network framework. The dashed line represent passing the hidden value at $l^{th}$ layer from encoder to decoder for denoising process.}
\label{fig:ladder}
\end{figure}

The ladder network is employed with both of the d-vector and x-vector frameworks. It is illustrated in Figure \ref{fig:ladder}, which consists of classification network and corresponding DAE. 

The classification network for classification task is a normal network (e.g. DNN for d-vector and TDNN for x-vector). The DAE for reconstruction task has two parts, encoder and decoder. The encoder shares the same parameters with the classification network, while the decoder tries to reconstruct the clean version of the noised input. Unlike conventional DAE, the ladder network adds Gaussian noise and performs denoising at each layer. Equation \ref{eq:noisyhidden} shows how to add the noise, where $\tilde{h}$ represents the noisy version of the hidden value, $l$ is the $l^{th}$ layer, and $\textit{N}$ is the Gaussian noise with zero mean and $\sigma$ standard deviation (we set $\sigma=0.3$ in this study). Equation \ref{eq:denoising} shows the denoising process relying on parameters of \ref{eq:mu} and \ref{eq:nu}, where the 10 parameters ($a_1$ to $a_{10}$) are learnable parameters, $u^{(l)}$ is batch normalized preactivation value propagated from $l+1^{th}$ layer in decoder, and $\hat{h}^{(l)}$ is the reconstructed hidden value at the $l^{th}$ layer in decoder. The denoising process performs reconstruction depending on two things, one is the batch normalized preactivation value from higher layer in decoder, and the other is the noisy version of the hidden value from the counterpart at the current layer in encoder.
\begin{equation}
  \tilde{h}^{(l)} = h^{(l)} + \textit{N}(0,\sigma)
\label{eq:noisyhidden}
\end{equation}
\begin{equation}
  \hat{h}^{(l)} = (\tilde{h}^{(l)}-\mu(u^{(l)}))\nu(u^{(l)})+\mu(u^{(l)})
\label{eq:denoising}
\end{equation}
\begin{equation}
  \mu(u^{(l)}) = a_{1}^{(l)}sigmoid(a_{2}^{(l)} u^{(l)}+a_{3}^{(l)})+a_{4}^{(l)}u^{(l)}+a_{5}^{(l)}
\label{eq:mu}
\end{equation}
\begin{equation}
  \nu(u^{(l)}) = a_{6}^{(l)}sigmoid(a_{7}^{(l)} u^{(l)}+a_{8}^{(l)})+a_{9}^{(l)}u^{(l)}+a_{10}^{(l)}
\label{eq:nu}
\end{equation}

The cost function for reconstruction is defined in Equation \ref{eq:daecost}, equal to the weighted summation of the square error between the reconstructed and original hidden values. The $\lambda^(l)$ is a scalar weighting the reconstruction loss at $l^{th}$ layer. In this study, we set $\lambda^{(0)}$ to 1000, $\lambda^{(1)}$ to 10 and $\lambda^{(l)}, l > 1$ to 0.1. It can be noticed that if we set $\lambda^{(l)} = 0, l >= 1$, the autoencoder is equivalent to a conventional DAE. 
\begin{equation}
\label{eq:daecost}
  denoising\;cost = \sum_{l}\lambda^{(l)}\left\|h^{(l)}-\hat{h}^{(l)}\right\|^2
\end{equation}

During training, the network will be updated with respect to the loss, which is the summation of the cross entropy loss of the SID task and the denoising cost. The proposed approach can therefore train a network distinguishing speakers' acoustic characteristics and maintain (and denoise) original input information as much as possible at each layer.

\begin{figure}[tb]
\centering
\subfigure[]
{
 \includegraphics[width=\columnwidth]{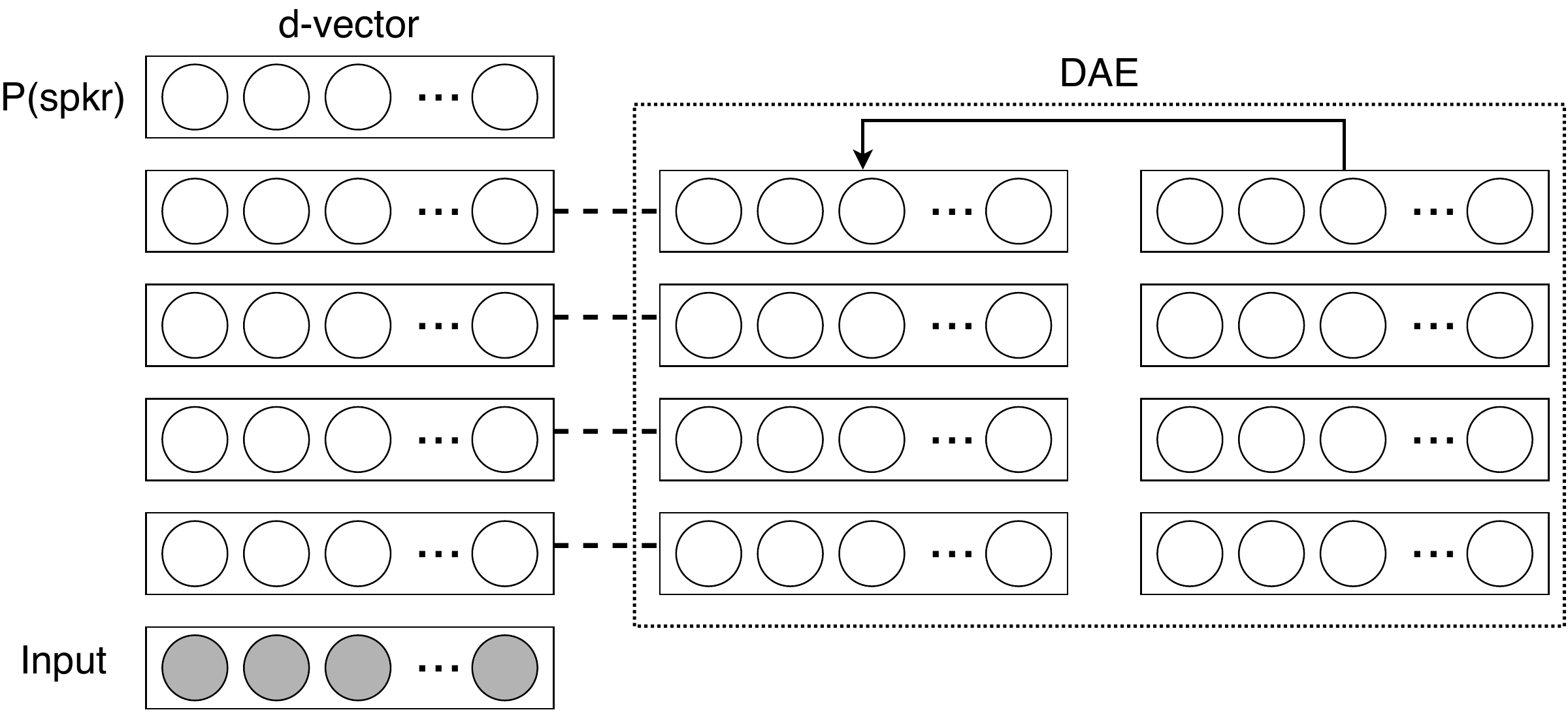}
\label{fig:d_ladder}
}
\subfigure[]
{
 \includegraphics[width=\columnwidth]{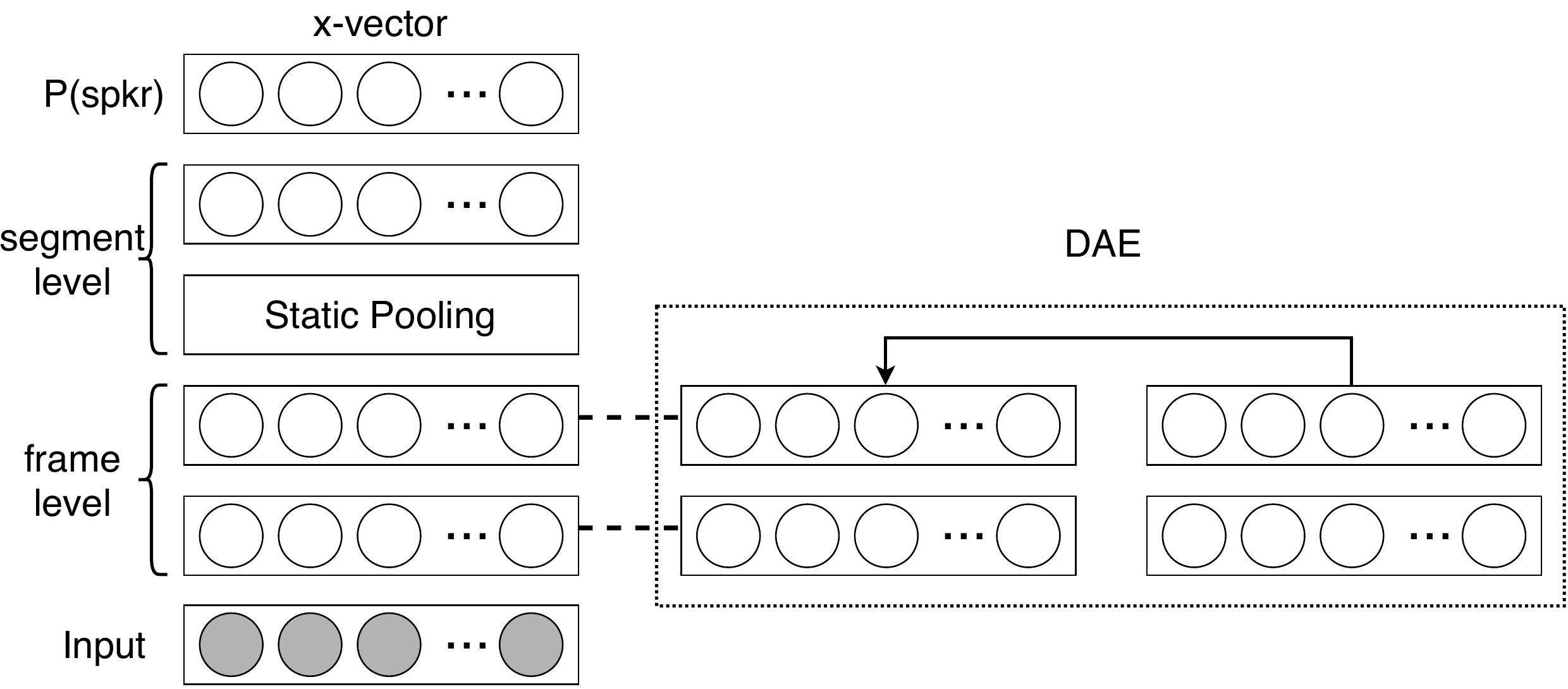}
\label{fig:x_ladder}
}\hspace{0.01mm}
\caption{(a) Diagram of the d-vector system with the ladder network. We name it as ``d-ladder". (b) Diagram of the x-vector system with the ladder network. The ladder framework is only applied on the frame-level layers. We name it as ``x-ladder". The dashed line between the DAE (denoising autoencoder) and the classification network (d-vector and x-vector in this case) represents the denoising cost. We did not plot the complete neural network for simplicity.}
\label{fig:sv_ladder}
\end{figure}

For d-vector framework, the classification network has 4 FC layers and 1 softmax layers. The encoder shares the weights of the FC layers with the classification network. The decoder has the mirrored architecture of the encoder. The proposed architecture (d-vector with ladder network) is plotted in Figure \ref{fig:d_ladder}.  For x-vector framework, the classification network has 9 layers (5 frame-level time-delay layers, 1 statistics pooling layer, 2 FC layers and 1 softmax layer). The encoder only shares the weights of the first 5 frame-level layers (all the layers before the statistics pooling layer), while the decoder has the mirrored architecture of the encoder. We did not do the reconstruction on the layers after the statistics layer because the statistics layer breaks the frame details and makes it impossible to reconstruct the original input segment.  The proposed architecture (x-vector with ladder network) is plotted in Figure \ref{fig:x_ladder}.

After training, we only use the classification network for embedding extractor. For d-vector, we extracted the hidden value from the layer right before the softmax layer. For x-vector, we extracted the hidden value from the first segment-level layer, which was the first FC layer right after the statistics pooling layer. This proposed approach did not introduce parameters compared with the baselines.

\section{Experiments and Results}
\label{sec:exper}

\subsection{Data and Features}
\label{sec:corpus}
Though we targeted at two practical use cases, we used public corpus for training and testing for fair comparison. We used VoxCeleb corpus (including Phase 1 and 2)~\cite{Nagrani17,Chung18b}, which is an audio-visual dataset consisting of short clips of human speech downloaded from YouTube. The human speech was mainly made during interviews, so the scenarios were natural and the background was not controlled. It includes 7363 speakers in total, where 39\% are females and 61\% are males. Each speaker spoke several segments of speeches. Duration of each segment varies from 3 to 20 seconds. The total duration is longer than 2000 hours. The corpus was originally separated into two sets "develop" and "test" in both phases. We used all the data from Phase 2 and the "develop" set from Phase 1 to train the model, and used "test" set from Phase 1 to evaluate the system, where there was 40 speakers. In this paper we focus on the speech based SV system, so we only used the speech part and ignored the videos.

We replicated the feature preparation described in \cite{variani2014deep} for d-vector framework and \cite{snyder2018x} for x-vector framework. A minor modification is made for d-vector is the input window was set to 51 frames, including 25 historic and 25 future frames. All features extraction and \emph{voice activity detection} (VAD) were performed by Kaldi~\cite{Povey_2011}.

\subsection{Experiment Settings}
For baselines, we used aforementioned d-vector \cite{variani2014deep} and x-vector \cite{snyder2018x} frameworks without ladder network, and name them as ``d-vector" and ``x-vector" in the following. We used different training approaches for the two approaches. For d-vector training, the optimizer was Adam \cite{Kingma_2014_2}. Each network was trained by running 15 epochs. After the first 5 epochs, the learning rate was halved every 2 epochs. For x-vector training, we replicated the Kaldi recipe in Tensorflow \cite{abadi2016tensorflow}. The training ran for 15 epochs so the results were comparable to the ones of the ``d-vector" systems. For the proposed approach, we use the same settings and training approaches, except we name them as ``d-ladder" and ``x-ladder" correspondingly. Across all the experiments, we did not use data augmentation. For final testing with the ladder networks, we only use the classification networks so that all the parameters number are the same.

\subsection{Experiment Results}
Table \ref{tab:summary} is the summary of the EER from the experiments, including the two baselines (``x-vector" and ``d-vector") and the proposed approach deployed on the baselines (``x-ladder" and ``d-ladder").

\begin{table}
\centering
\begin{tabular}{c|c}
\hline
Approach & EER [\%]\\
\hline
\hline
d-vector & 20.1 \\
d-ladder & \textbf{18.1} \\
\hline
x-vector & 3.86 \\
x-ladder & \textbf{3.78} \\
\hline
\end{tabular}
\caption{Summary of the experiment result. The ``EER'' is the equal-error-rate. The font in bold stands for the best performance in corresponding categories.}
\label{tab:summary}
\end{table}

The results show that the ``x-vector'' framework can significantly outperform the ``d-vector'' framework. The gain was mainly from the long-term information on the segment level, because ``d-vector'' only took around 0.5 sec input, while ``x-vector'' was modeled on the segments longer than 2 secs.

The proposed approach can outperform the baseline in both cases. The ``d-ladder" outperformed the ``d-vector" by 10\% relative difference, while the ``x-ladder" outperformed the ``x-vector" by 2\% relative difference. The improvement in the ``x-vector" framework was smaller, because the baseline, ``x-vector" has already been well developed. There was limited space for improvement. It should be noticed that the improvement was from no extra data, no augmentation and no extra parameters.

The ladder network looks similar to multi-task learning with the secondary task of reconstructing the original input. To verify the capability of the ladder network further, we compared the proposed framework with the multi-task learning approach. The evaluating result is shown in Table \ref{tab:multi_task}, where the ``x-multi" approach stands for the x-vector with multi-task learning. We performed the comparison under the x-vector framework, because this was the framework with the best performance in our study. The x-multi consisted of two tasks, speaker identification and original input reconstruction. The x-multi architecture was similar to the ``x-ladder" (shown in Figure \ref{fig:x_ladder}). The speaker identification task was performed on top the of the x-vector network, while the original input reconstruction was performed on the frame-level layers. We built a DAE for the frame-level layers, and kept its parameters number same as the DAE's in x-ladder. The results show that the ``x-ladder" can outperformed the x-multi, indicating the capability of the proposed approach. It should be noticed that the x-multi had worse performance than the x-vector, which means brutally adding reconstruction task to the network for speaker identification may not be helpful. It can be explained that the regularization from the reconstruction task was not imposed much on to the frame-level layers, because the error back-propagated from the reconstruction task may mainly apply on the tuning of the decoding side in DAE (the mirrored part of the frame-level layers). In contrast, the ladder network forced each layer in the classification network to be tuned regarding with the reconstruction errors, which was a strong regularization.

\begin{figure}[tb]
\centering
\subfigure
{
 \includegraphics[width=\columnwidth]{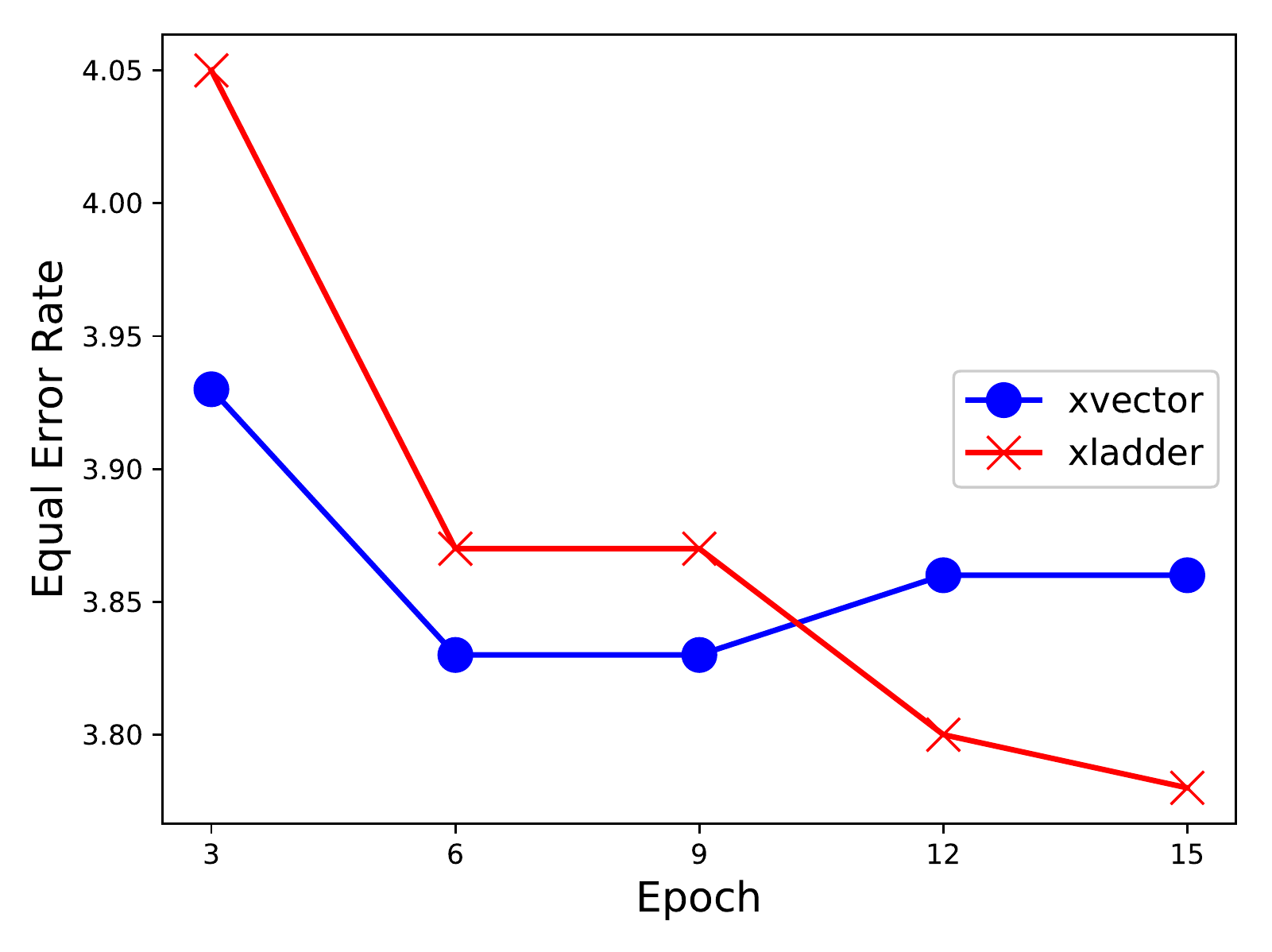}
}
\caption{The EER (evaluated on testing data) every 3 epochs from the intermediate models in the training of the ``x-vector" and ``x-ladder" systems. The performance at the 15th epoch was reported as the final result.}
\label{fig:xv_results}
\end{figure}

We also observe that the training process for the ladder network took longer. The training time of the ``d-vector" system was about 40 hrs on the Nvidia RTX 2080 ti, while the ``d-ladder" system took 45 hrs. The training time of the ``x-vector" system was about 106 hrs, while the ``x-ladder" system took 180 hrs. We analyzed the training process for the ``x-vector" framework in details (because it was the best system in our study). We evaluated the performance from the intermediate models (models per 3 epochs) during training and plot the results in Figure \ref{fig:xv_results}. It can be seen that the performance curve of the ``x-vector" system vibrated along the epochs, indicating it had over-fitting problem. But the curve of the ``x-ladder" system was more stable and always went down. It implies the system had better generalization, since the results did not show any clue of over-fitting. We did not continue the training process, because we set 15 epochs as a global limit for all systems' training. We could expect better performance when the training continued. Future work is needed to investigate how much improvement the ``x-ladder" system can achieve. This proved that the ladder network can improve the system generalization in the speaker verification task. 

\begin{table}
\centering
\begin{tabular}{c|c}
\hline
Approach & EER [\%]\\
\hline
\hline
x-vector & 3.86 \\
x-multi & 4.25 \\
x-ladder & \textbf{3.78} \\
\hline
\end{tabular}
\caption{Comparison between ladder network and multi-task learning. ``x-multi" stands for x-vector framework with multi-task learning; ``x-ladder" stands for x-vector framework with the ladder network approach.}
\label{tab:multi_task}
\end{table}

\section{Conclusions}
\label{sec:concl}
In this paper, we propose to apply a new approach to train the neural network for extracting better speaker embeddings. The proposed approach consisted of a modified DAE and a feed forward classification network. During training, the supervised and unsupervised learning were both applied to balance keeping and discarding information. The unsupervised learning helped the network not to bias to the task in the training data domain, therefore it is especially suitable for the speaker verification problem which is an open set problem. We evaluate our approach on Voxceleb corpus with d-vector and x-vector frameworks, which are the state-of-the-art approaches respectively tackling with short and long duration audios. The results showed that our approach can have at most 10\% relative improvement without data augmentation and extra parameters, which proves the advantage of the proposed approach. We also further analyzed the ladder network under the x-vector framework. The analysis indicates the ladder network can help regularize and generalize the speaker identification network.


\bibliographystyle{IEEEtran}

\bibliography{reference}

\end{document}